\documentclass[prl,aps,twocolumn,showpacs,amsmath,amssymb,floatfix,superscriptaddress,citeautoscript]{revtex4}
\usepackage{graphicx}
\usepackage{dcolumn}
\usepackage{bm}
\usepackage{amsmath}
\usepackage{amssymb}
\usepackage[normal]{subfigure}

 \makeatletter

\usepackage{color}

\begin{document}

\title{Floquet Majorana Fermions for Topological Qubits}

\author{Dong E. Liu}\email{delphy@msu.edu}
\affiliation{Department of Physics and Astronomy, Michigan State
University, East Lansing, Michigan 48824, USA}

\author{Alex Levchenko}
\affiliation{Department of Physics and Astronomy, Michigan State
University, East Lansing, Michigan 48824, USA}

\author{Harold U. Baranger}
\affiliation{Department of Physics, Duke University, Box 90305,
Durham, North Carolina 27708, USA}

\date{November 9, 2012}

\begin{abstract}
We introduce and develop an approach to realizing a topological
phase transition and non-Abelian statistics with dynamically induced
Floquet Majorana fermions (FMFs). When the periodic driving
potential does not break fermion parity conservation, FMFs can
encode quantum information. Quasi-energy analysis shows that a
stable FMF zero mode and two other satellite modes exist in a wide
parameter space with large quasi-energy gaps, which prevents
transitions to other Floquet states under adiabatic driving. We also
show that in the asymptotic limit FMFs preserve non-Abelian
statistics and, thus, behave like their equilibrium counterparts.
\end{abstract}


\maketitle

\textit{Introduction} --- Proposals of solid
state~\cite{moore91,read00,fu08,sau10,aliceaPRB10,Lutchyn10,oreg10}
and cold atomic~\cite{sato09,zhu11,jiang11} systems hosting Majorana
fermions (MFs) have been a recent focus of attention. These systems
present novel prospects for quantum computation since a widely
separated pair of MF bound states, that formally correspond to
zero-energy states of an effective Bogoliubov-de Gennes (BdG)
Hamiltonian, forms a nonlocal fermionic state that is immune to
local sources of decoherence. Moreover, MFs obey non-Abelian
statistics and thus have potential for topological quantum
information processing. Among the key signatures of MFs are a
zero-bias resonance in tunneling~\cite{law09,sauRPB10}, half-integer
conductance quantization~\cite{wimmer11,liu11}, and a $4\pi$
Josephson effect~\cite{kitaev}. Some of these predictions have
already received possible experimental
support~\cite{kouwenhovenSCI12,ggordon12,rokhinson12}.

Topological states of matter can be induced dynamically by
time-periodic driving, the so-called Floquet
approach~\cite{inoue10,kitagawa10,lindner11}. This brought to the
agenda the new concept of Floquet Majorana fermion
(FMFs)~\cite{jiang11}. It turns out that even if the system is
initially in the topologically trivial state, its Floquet version
may exhibit topological properties. A realization of such states
where they can be readily manipulated and precisely tuned in a wide
parameter space is therefore highly desirable. The natural questions
for FMF systems are: whether they are robust and tunable, whether
they can encode quantum information, and whether they follow
non-Abelian statistics as for their equilibrium counterparts. Our
study aims to answer these questions.

We consider a generic platform to investigate non-Abelian statistics
and potentially to realize topological quantum computation based on
FMFs. The model is broadly applicable to both
semiconductor-superconductors heterostructures with strong
spin-orbit interaction and in-plane magnetic
field~\cite{Lutchyn10,oreg10}, and to cold atomic systems where
superconducting order is controlled by Feshbach resonances while
spin-orbit coupling and Zeeman field effects are induced by an
optical Raman transition~\cite{jiang11}. The latter realization is
practically more promising since it allows a greater degree of
control. Furthermore, atomic condensates can be isolated thus
suppressing dissipation on long time scales.

We show, first, that if FMFs exist, they will exist at any
instantaneous time. Therefore, FMFs can encode quantum information
if the driving potential does not break fermion parity conservation.
We study the quasi-energy spectrum of the problem analytically by
using a rotating frame analysis in the limit that the frequency is
large compared to the band width. We also perform exact numerical
calculations which capture certain features of the spectrum beyond
the rotating wave approximation (RWA). A broad range of parameters
supporting FMFs is identified as a function of driving frequency
$\omega$ and amplitude $K$ for three specific driving scenarios:
periodic modulation of the chemical potential, or the Zeeman field. Finally, by using a two-time
formalism~\cite{breuer89,althorpe97}, we show that FMFs follow the
same non-Abelian statistics as their stationary counterparts. This
conclusion stems from the observation that a generalized Floquet
Berry matrix does not affect the non-Abelian statistics of FMFs
since large quasi-energy gap ensures no transitions to other Floquet
quasi-energy states in the adiabatic movement.

{\em Floquet Theorem for Majorana Fermion ---} Let us consider
Floquet theory~\cite{shirley65}. Suppose that the Hamiltonian has an
explicit time dependence $\hat{H}(t)=\hat{H}(t+T)$ with period
$T=2\pi/\omega$, where $\omega$ is the driving frequency. The
solution of the Schr\"{o}dinger equation can be described by a
complete set of time-dependent state
$|\Phi_{\alpha}(t)\rangle=e^{-i\epsilon_{\alpha}t}|\phi_{\alpha}(t)\rangle$,
where quasi-energies $\epsilon_{\alpha}$ satisfy the equation
$[\hat{H}(t)-i\partial_t]|\phi_{\alpha}(t)\rangle =
\epsilon_{\alpha} |\phi_{\alpha}(t)\rangle$ and
$|\phi_{\alpha}(t)\rangle=|\phi_{\alpha}(t+T)\rangle$ are Floquet
states (hereafter $\hbar=1$). The evolution operator
$\hat{U}(t)=\mathbb{T} \exp(-i\int^{T}_{0} \hat{H}(t) dt)$ has the
following property
\begin{equation}
\hat{U}(t+T,t)|\phi_{\alpha}(t)\rangle =
e^{-i\epsilon_{\alpha}T}|\phi_{\alpha}(t)\rangle.
\label{eq:evolution}
\end{equation}
One can define an effective Hamiltonian $\hat{H}_{\rm eff}(t)$
through the relation~\cite{kitagawa10,lindner11}
\begin{equation}
 \hat{U}(t+T,t)\equiv e^{-i \hat{H}_{\rm eff}(t) T},
\end{equation}
with $\hat{H}_{\rm eff}(t)|\phi_{\alpha}(t)\rangle=
\epsilon_{\alpha}|\phi_{\alpha}(t)\rangle$. We treat $t$ as just a
parameter. The effective Floquet Hamiltonian is defined at each
instantaneous time, and the topological properties of each of these Hamiltonians is the same
\cite{kitagawa10,lindner11}.

If the system is described by a BdG Hamiltonian, the quasi-particle
excitation spectrum will possess a particle-hole symmetry even if
the time-dependent potential is added \cite{kitagawa10}. For any
quasi-energy state
$|\phi_{\epsilon}(t)\rangle=\hat{\gamma}_{\epsilon}(t)|0\rangle$,
the relation
$\hat{\gamma}_{\epsilon}(t)=\hat{\gamma}_{-\epsilon}^{\dagger}(t)$
is guaranteed. So, the zero quasi-energy state reveals the existence
of a Floquet MF \cite{jiang11}. The full wavefunction for
$\epsilon_{0}=0$ can be written as $|\Phi_{0}(t)\rangle =
e^{-i\epsilon_{0}}|\phi_{0}(t)\rangle = |\phi_{0}(t)\rangle =
\hat{\gamma}_{0}(t)|0\rangle$, with
$\hat{\gamma}_{0}(t)=\hat{\gamma}_{0}^{\dagger}(t)$. Since
quasi-energy is only defined in a interval of $\omega$ (e.g. from
$-\omega/2$ to $\omega/2$), another type of Floquet MF exists at
$\epsilon=\pm\omega/2$ with $e^{-i\omega
t/2}\gamma_{\omega/2}=[e^{-i\omega
t/2}\hat{\gamma}_{\omega/2}]^{\dagger}$~\cite{jiang11}. From the
argument above, we can show that if the zero quasi-energy state
exists, a zero energy Floquet MF mode $\gamma(t)$ exists at any
instantaneous time $t$. The MF operator evolves in time periodically
$\hat{\gamma}(t)=\hat{\gamma}(t+T)$; in general, it is different at
different instantaneous times,
$\hat{\gamma}(t)\neq\hat{\gamma}(t')$.

{\em Quasi-Energy Spectrum and Floquet Majorana Fermion ---} To
demonstrate the existence of FMFs consider a one dimensional wire
with Rashba spin-orbit interaction $\lambda_{\rm SO}$, Zeeman splitting
$V_z$, and proximity-induced superconducting term $\Delta$. The
system can be described by a tight-binding
Hamiltonian~\cite{Lutchyn10,oreg10,jiang11}:
\begin{eqnarray}
\hat{H}_{0} & =& \sum_{i,\sigma}
\left[-\eta\left(\hat{c}_{i+1\sigma}^{\dagger}\hat{c}_{i\sigma}+h.c.\right)+
\mu\hat{c}_{i\sigma}^{\dagger}\hat{c}_{i\sigma}\right]\nonumber \\
 &+&\sum_{i}V_{z}
 \left(\hat{c}_{i\uparrow}^{\dagger}\hat{c}_{i\uparrow}-\hat{c}_{i\downarrow}^{\dagger}
 \hat{c}_{i\downarrow}\right)
 +\Delta\sum_{i}\left(\hat{c}_{i\uparrow}^{\dagger}\hat{c}_{i\downarrow}^{\dagger}+h.c.\right) \nonumber \\
 &+&\lambda_{\rm SO}\sum_{i}
 \left(\hat{c}_{i+1\uparrow}^{\dagger}\hat{c}_{i\downarrow}-
 \hat{c}_{i+1\downarrow}^{\dagger}\hat{c}_{i\uparrow}+h.c.\right),
\label{eq:H0}
\end{eqnarray}
Here, $i$ and $\sigma=\uparrow\downarrow$ denote fermion site and
spin indices while $\hat{c}_{i\sigma}(\hat{c}^\dag_{i\sigma})$ are
corresponding operators, $\eta$ is the hopping term along the chain
which yields a band width $D=4\eta$, and $\mu$ is the chemical
potential of the lattice model which is set to the particle-hole
symmetric point~\cite{note}. Note that Hamiltonian Eq.~\eqref{eq:H0}
is equally generic for a system of cold atoms~\cite{jiang11}.

To add time dependence, it is natural to consider modulating one of
the parameters in $\hat H_0$: the chemical potential, the spin-orbit
coupling \cite{reynoso12}, or the Zeeman field. We first consider
periodic modulation of the chemical potential; the Hamiltonian is
$\hat{H}(t)=\hat{H}_{0}+\hat{H}_\mu(t)$ with
\begin{equation}
 \hat{H}_\mu(t)= K \cos(\omega t) \sum_{i} (\hat{n}_{i\uparrow}+\hat{n}_{i\downarrow}),
\end{equation}
where
$\hat{n}_{i\sigma}=\hat{c}_{i\sigma}^{\dagger}\hat{c}_{i\sigma}$. To
calculate the quasi-energy, one can choose a basis in the rotating
frame~\cite{eckardt05}
\begin{equation}
 | \{n_{i\sigma}\};m\rangle = e^{ -\frac{iK\sin(\omega t) }{\omega}
\sum_{i} (\hat{n}_{i\uparrow}+\hat{n}_{i\downarrow}) +i m\omega t} \;| \{n_{i\sigma}\}\rangle\; ,
\end{equation}
where $|\{n_{i\sigma}\}\rangle$ is the basis of the unperturbed
system, and $m$ labels the photon sector of the Floquet basis. The
quasi-energy can be obtained by diagonalizing the Floquet operator
$\hat{H}(t)-i\partial_t$ in this basis. The orthonormality condition
of the Floquet states is only defined in an extended Hilbert
space~\cite{sambe73}, so the inner product must include an extra
time integral over a full period: $\langle\langle\cdot
|\cdot\rangle\rangle=(1/T)\int_{0}^{T} dt
\langle\cdot|\cdot\rangle$. The matrix elements read
\begin{eqnarray}
 & &\langle\langle \{n_{i\sigma}\};m |\hat{H}(t)-i\partial_t |\{n_{i\sigma}'\};m'  \rangle\rangle \nonumber\\
 & =& \frac{1}{T}\int_{0}^{T} dt \langle\{n_{i\sigma}\}|
 e^{\frac{iK\sin(\omega t)}{\omega}\sum_{i} (\hat{n}_{i\uparrow}+\hat{n}_{i\downarrow})}
   \left(\hat{H}_0 + m\omega \right) \nonumber\\
 & & \times e^{-\frac{iK\sin(\omega t)}{\omega}\sum_{i} (\hat{n}_{i\uparrow}+\hat{n}_{i\downarrow})}
    |\{n_{i\sigma}'\}  \rangle e^{-i(m-m')\omega t}.
\label{eq:FME}
\end{eqnarray}
Since different photon sectors are separated by an energy gap of
order $\omega$, in the limit $\omega\gg D$, the admixture of photon
sectors can be neglected; this is in essence the rotating wave
approximation. Then, we can consider only the zero photon sector and
obtain an effective Floquet Hamiltonian by computing the $m=m'=0$
matrix element. The key point to notice is that only the
superconducting term in \eqref{eq:H0} fails to commute with the
chemical potential operator $\sum_{i}
(\hat{n}_{i\uparrow}+\hat{n}_{i\downarrow})$. Evaluation of
Eq.~(\ref{eq:FME}) within the RWA yields an effective Floquet
Hamiltonian with exactly the same form as $\hat{H}_0$ with the
pairing coupling $\Delta$  effectively renormalized to
\begin{equation}\label{eq:Delta-eff}
\Delta_{\mathrm{eff}}=\Delta J_0(2K/\omega).
\end{equation}
($J_{0}(x)$ is the zero order Bessel function of the first
kind.)

We conclude from Eq.~\eqref{eq:Delta-eff} that in Floquet systems
one may induce a topological phase transition dynamically. Indeed,
recall that the regime for a topological superconducting phase of
$\hat{H}_{0}$, which supports MFs, requires the condition
$V_z^2>\Delta^2+(\mu+2\eta)^2$ ~\cite{Lutchyn10,oreg10,note}. Even
if initially this condition is not satisfied so that the system is
in the topologically trivial state, the renormalization
$\Delta\to\Delta_{\mathrm{eff}}$ may make a topological phase
possible since $\Delta_{\mathrm{eff}}<\Delta$. Thus, periodic
modulation of the chemical potential provides a way to tune the
topological phase and so realize MFs by varying the parameter
$K/\omega$. The rescaling Eq.~\eqref{eq:Delta-eff} holds only, of
course, to the extent that off-diagonal couplings can be neglected;
we address the generic case numerically below and show that more
dramatic changes in behavior are entirely possible.

For periodic modulation of the Zeeman field, a similar analysis can
be carried out by adding $\hat{H}_z(t)= K \cos(\omega t) \sum_{i}
(\hat{n}_{i\uparrow}-\hat{n}_{i\downarrow})$ to $\hat H_0$. Since
only the Rashba term in Eq.~(\ref{eq:H0}) does not commute with the
Zeeman term, the spin-orbit parameter is modified in the effective
Floquet Hamiltonian: $\lambda_{\rm SO}\rightarrow \lambda_{\rm
SO}J_{0}(2K/\omega)$. Thus, periodic Zeeman modulation cannot induce
a topological phase transition if one keeps only the zero photon
sector. However, numerical investigation beyond the RWA [keeping all
off-diagonal blocks of the effective Floquet Hamiltonian $\propto
J_{m-m'}(2K/\omega)$] reveals that FMFs do, in fact, appear, and so
we now turn to our numerical results.

\begin{figure}[t]
\centering
\includegraphics[width=3.2in,clip]{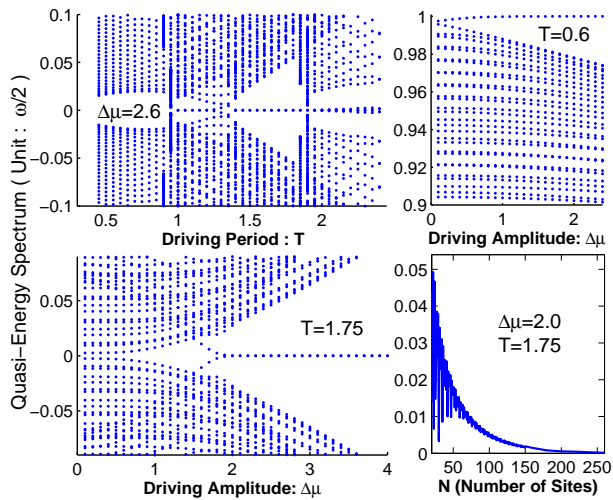}
\vspace{-0.2in} \caption{(color online) Quasi-energy spectrum for
square-wave driven chemical potential. Parameters: $\eta=1.5$ (full
band-width $D=4\eta=6.0$), $V_z=1.0$, $\Delta=1.0$, $\lambda_{\rm
SO}=1.2$, and $(\mu_1+\mu_2)/2=0.5$. [Left panels]: quasi-energy near
$\epsilon=0$, as a function of driving period $T$ for
$\Delta\mu=|\mu_1-\mu_2|=2.6$ (upper), and as a function of driving
amplitude $\Delta\mu$ for $T=1.75$ (lower). [Right upper panel]:
quasi-energy near $\epsilon=\omega/2$ as a function of driving
amplitude $\Delta\mu$ for $T=0.6$. [Right lower panel]: Finite size
splitting (indicating the coupling between two FMFs at the two ends) for
$\epsilon=0$ mode as a function of the number of sites in the
chain ($T=1.75$, $\Delta\mu=2.0$). The finite size splitting shows
exponential suppression accompanied by oscillations. There are
$N=260$ sites in the chain. Note: the unit used for the quasi-energies is
$\omega/2=\pi/T$. } \label{fig:DrivenMu}
\end{figure}

For numerical convenience we consider square-wave driving of the
chemical potential or Zeeman field: $\mu =\mu_1$ for
$nT<t<(n+1/2)T$, and $\mu=\mu_2$ for $(n+1/2)T<t<(n+1)T$ (with
$n=0,1,2,...$), and similarly for $V_z$. The evolution operator for
the full period reads then $\hat{U}(T,0)=$
$e^{-i\frac{\hat{H}_{2}T}{2\hbar}}e^{-i\frac{\hat{H}_{1}T}{2\hbar}}$,
and the quasi-energy spectrum $\epsilon_{\alpha}$ is obtained
numerically using Eq.~(\ref{eq:evolution}). In all cases here, the
parameters at any instantaneous time correspond in the static system
to the topologically trivial phase.

The results for periodically modulated chemical potential are shown
in Fig.~\ref{fig:DrivenMu}. Clearly, one obtains stable $\epsilon=0$
Floquet Majorana zero modes (left panels) for a large range of
parameters, as well as $\epsilon=\omega/2$ modes (upper right panel)
\cite{supp}. Note that the parameters used in
Fig.~\ref{fig:DrivenMu} are very far from those for which the RWA
result Eq.~\eqref{eq:Delta-eff} yields a FMF: here $V_z^2 -
(\mu+2\eta)^2 < 0$ at all times, so no renormalized $\Delta$ can
yield a non-trivial phase. Nevertheless, FMF appear once $\Delta\mu$
surpasses a threshold $\Delta\mu_c$. The figure shows that the
threshold for an $\epsilon=\omega/2$ FMF can be very small compared
to that for an $\epsilon=0$ FMF, and also that the quasi-energy gap
can be tuned by varying $\Delta\mu$. The splitting of a $\epsilon=0$
mode due to finite size effects is plotted in the right lower panel;
it shows the expected decay of the level splitting as the number of
sites, and hence the separation between the two FMF, increases.

\begin{figure}[t]
\centering
\includegraphics[width=3.4in,clip]{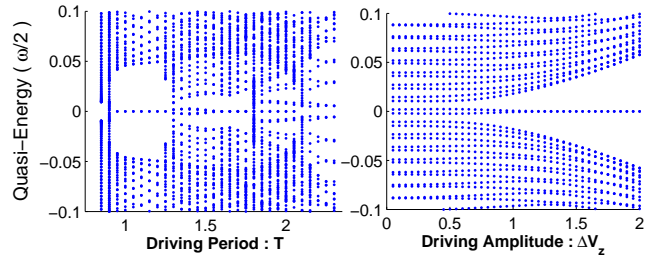}
\vspace{-0.22in}
\caption{(color online) Quasi-energy spectrum for square-wave
driving of the Zeeman splitting, near $\epsilon=0$. Parameters: $\eta=1.4$ (full
band-width $D=4\eta=5.6$), $V_z=1.0$, $\Delta=2.0$,
$\lambda_{\rm SO}=1.5$, and $(\mu_1+\mu_2)/2=0.0$. Left panel:
quasi-energy as a function of driving period $T$,
for $\Delta V_z=|V_{z1}-V_{z2}|=1.8$. Right panel:
quasi-energy as a function of driving amplitude $\Delta V_z$, for $T=1.1$.
There are 260 sites in the chain.}
\label{fig:DrivenMF}
\end{figure}

The quasi-energy spectrum with periodic Zeeman splitting is shown in
Fig.~\ref{fig:DrivenMF}. It also reveals FMFs. We stress once
again that to obtain FMF in this case, the RWA is not enough and
off-diagonal blocks of the Floquet Hamiltonian are crucial.

{\em  Floquet Topological Qubit and Non-Abelian Statistics ---} A
natural question is whether FMFs can form topological qubits, as
their static counterparts do. FMF can certainly encode quantum
information: an FMF exists at all instantaneous times, and neither
chemical potential driving nor Zeeman driving changes the total
fermion parity. Then, a more difficult question is whether FMFs obey
non-Abelian statistics. We will provide an argument for a 2D system,
which can then be generalized to a 1D network following the argument
for static MF \cite{alicea11}.

Suppose that FMFs are moved (which can be achieved by tuning the
driving potential on and off, or changing the driving amplitude,)
along a path $R(t)$ with the Schr\"{o}dinger equation
$[\hat{H}(R(t),t)-i\partial_t]|\Phi(t)\rangle=0$. The position of the FMF
$R(t)$ is assumed to vary on a very slow time scale compared to the
fast periodic driving. Then, it is convenient to separate the fast
and slow time scales, and apply the two-time formalism of Floquet
theory~\cite{breuer89,althorpe97}: $i\partial_t\rightarrow
i\partial_t + i\partial_\tau$, where $t$ indicates the fast time and
$\tau$ denotes the slow time. Then the Schr\"{o}dinger equation
becomes
\begin{equation}
 i\partial_{\tau}|\Phi(R(\tau),t)\rangle =
 \left[\hat{H}(R(\tau),t)-i\partial_t  \right] |\Phi(R(\tau),t)\rangle.
\end{equation}
When we consider the dynamics on the slow time scale, the fast time
$t$ can be considered as a parameter. It was pointed out by Breuer
and Holthaus \cite{breuer89} (also see a recent discussion
\cite{kitagawaGF11}) that a Floquet system follows a generalized
adiabatic theorem. Define the instantaneous (for $\tau$)
quasi-energy eigenstates using the Floquet operator
\begin{equation}
 \left[\hat{H}(R(\tau),t)-i\partial_t  \right]
 |\phi_{\alpha}(R(\tau),t)\rangle = \epsilon_{\alpha}(R(\tau))|\phi_{\alpha}(R(\tau),t)\rangle .
\end{equation}
Suppose the system is initially in a Floquet state
$|\Phi(R(\tau=0),t)\rangle=|\phi_{\alpha}(R(\tau=0),t)\rangle$.
Standard procedures in quantum mechanics can be applied to Floquet
states as long as the extended inner product mentioned above,
$\langle\langle\cdot|\cdot\rangle\rangle$, is used. Second order
perturbation theory then yields \cite{breuer89,kitagawaGF11}
\begin{eqnarray}
 &|\Phi(R(\tau),t)\rangle = e^{-i\theta_{\alpha}(\tau)} e^{-i\chi_{\alpha}(\tau)}
  \Big( |\phi_{\alpha}(R(\tau),t)\rangle \qquad \nonumber\\
 & - \sum_{\beta\neq\alpha}|\phi_{\beta}(R(\tau),t)\rangle
   \frac{\langle\langle \phi_{\alpha}(R(\tau))
   | i\partial_{\tau}
   |\phi_{\beta}(R(\tau))\rangle\rangle}{\epsilon_{\beta}(R(\tau))-\epsilon_{\alpha}(R(\tau))}
   \Big) \;,
   \label{eq:berry}
\end{eqnarray}
where
$\theta_{\alpha}(\tau)=\int_{0}^{\tau}d\tau'\epsilon_{\alpha}(R(\tau'))$
is the dynamical phase, and
$\chi_{\alpha}(\tau)=\int_{0}^{\tau}d\tau'\langle\langle
\phi_{\alpha}(R(\tau')) | i\partial_{\tau'}
|\phi_{\alpha}(R(\tau'))\rangle\rangle$ is the generalized Berry
phase. Therefore, to avoid transitions to other quasi-energy states,
the change in time scale $\tau$ must be slow and the quasi-energy
gap should be large:
$|\epsilon_{\beta}(R(\tau))-\epsilon_{\alpha}(R(\tau))|
\gg|\langle\langle \phi_{\alpha}(R(\tau))| i\partial_{\tau}
|\phi_{\beta}(R(\tau))\rangle\rangle|$. We assume this condition is
satisfied so that the system will stay in its initial Floquet state.

The Floquet Majorana excitations can be described by a Bogoliubov
quasi-particle operator,
\begin{equation}
 \hat{\gamma}^{\dagger}(t) =\!
 \int \!d\textbf{r} \big[ u(\textbf{r},R(\tau),t)\hat{\psi}^{\dagger}(\textbf{r})
     +  v(\textbf{r},R(\tau),t)\hat{\psi}(\textbf{r}) \big],
\label{eq:bogoliubov}
\end{equation}
where $\hat{\psi}^{\dagger}(\textbf{r})$ ($\hat{\psi}(\textbf{r})$)
creates (annihilates) a fermion at $\textbf{r}$, and $v=u^{*}$ for a
MF. A $U(1)$ gauge transformation which changes the superconducting
order parameter phase by $2\pi$ \cite{ivanov01} is allowed by using
the extended space of the Floquet system \cite{supp}. This causes a
minus sign on both $\hat{\psi}^{\dagger}(\textbf{r})$ and
$\hat{\psi}(\textbf{r})$, changing the sign of the FMF operator as
well. Due to such multivaluedness, a branch cut is necessary to
define the phase of the wave function. So, the exchange of two FMFs
$\hat{\gamma}_i(t)$ and $\hat{\gamma}_j(t)$ can induce a
transformation: $\hat{\gamma}_i(t)\rightarrow\hat{\gamma}_j(t)$ and
$\hat{\gamma}_j(t)\rightarrow -\hat{\gamma}_i(t)$ (since one of the
FMF, say $\hat{\gamma}_j(t)$, must pass through the branch cut). For
a 1D network, the exchange of two FMFs (through a T-junction, for
instance) flips the sign of the superconducting pairing term, which
results in exactly the same transformation as in the 2D case
\cite{alicea11}.

Given two FMFs $\hat{\gamma}_1(t)$ and $\hat{\gamma}_2(t)$, one can
form a non-local regular fermion
$\hat{d}^{\dagger}(t)=(\hat{\gamma}_1(t)+i\hat{\gamma}_2(t))/\sqrt{2}$.
Let $|G(t)\rangle$ be the Floquet BCS state which is annihilated by
any Floquet quasi-energy operators. $|G(t)\rangle$ and
$\hat{d}^{\dagger}(t)|G(t)\rangle$ form a two-fold degenerate space.
The exchange of two MFs results in $|G(t)\rangle\rightarrow
e^{i\varphi}|G(t)\rangle$ and
$\hat{d}^{\dagger}(t)|G(t)\rangle\rightarrow
e^{i\varphi}e^{i\pi/2}\hat{d}^{\dagger}(t)|G(t)\rangle$. The $\pi/2$
phase difference after the transformation signifies non-Abelian
statistics \cite{stone06,cheng11}.

The exchange of two MF can also induce an extra unitary evolution
involving a non-Abelian Berry matrix \cite{wilczek&zee}. The form of
the matrix can be generalized to a Floquet system~\cite{supp} by
replacing $\langle\cdot|\cdot\rangle$ with
$\langle\langle\cdot|\cdot\rangle\rangle$; the unitary evolution
then reads
\begin{equation}
 \hat{U}(\tau)=\mathbb{P}\exp\left[ i\int_0^{\tau} \textbf{M}(\tau') d\tau' \right]
\end{equation}
where $\mathbb{P}$ denotes path-ordering and $\textbf{M}_{\alpha\beta}(\tau)=\langle\langle
\phi_{\alpha}(R(\tau))| i\partial_{\tau}
|\phi_{\beta}(R(\tau))\rangle\rangle$ is the generalized non-Abelian
Berry matrix ~\cite{supp}. We want to test whether $\mathbf{M}_{\alpha\beta}$
causes any extra phase difference that breaks the non-Abelian
statistics of FMFs. First, the non-diagonal matrix elements of
$\mathbf{M}_{\alpha\beta}$ are zero since fermion parity is
conserved (as emphasized above this is true for all driving
scenarios). Second, we follow a procedure similar to that for a stationary MF
\cite{stone06,cheng11} where the odd parity element $i\langle\langle
G|\hat{d}\;\partial_{\tau}\,\big(\hat{d}^{\dagger}|
G\rangle\rangle\big)$ is written as the sum of the even parity
element $i\langle\langle G|\partial_{\tau}| G\rangle\rangle$ and an
extra term $i\langle\langle G| ( \hat{d}
\partial_{\tau}\hat{d}^{\dagger} )| G\rangle\rangle$. It is just
this term that might affect the the phase difference $\pi/2$ and so
the non-Abelian statistics. By using Eq.~(\ref{eq:bogoliubov}) and
the MF condition $v_i=u_{i}^{*}$ one finds
\begin{equation}
 \langle\langle G| ( \hat{d}
\partial_{\tau}\hat{d}^{\dagger} )| G\rangle\rangle=
\frac{2i}{T}\int_0^T\!\! dt\!\! \int d\textbf{r}
 \rm{Re}(u_1^{*}\partial_{\tau}u_2 - u_2^{*}\partial_{\tau}u_1).
\end{equation}
This term is exponentially small since it contains an overlap of
wave functions for spatially separated MFs; furthermore, it actually
vanishes in the adiabatic limit. We conclude that the non-Abelian Berry
phase does not destroy the desired statistics of FMFs.

\textit{Summary} --- Periodic modulation of the chemical potential
or the Zeeman field appears to be a promising way to produce FMFs,
both of which can be realized in 1D cold atom condensates. We find
that Floquet MFs are robust and can be generated in a wide parameter
range. This system may have potential for topological quantum
computation since FMFs obey the same non-Abelian statistics as their
equilibrium counterparts.

D.E.L and A.L. acknowledge support from Michigan State University.
The work at Duke was supported by US DOE, Division of Materials
Sciences and Engineering, under Grant No.\,DE-SC0005237.

\begin{widetext}

\global\long\def\theequation{S\arabic{equation}}
\global\long\def\thefigure{S\arabic{figure}}

\section{Supplementary Information}

In this supplementary information we (i) provide more details about
the rotating frame analysis (specifically, the derivation of Eq. (7)
in the main text), (ii) provide more data on the quasi-energy
spectrum, (iii) discuss $U(1)$ gauge invariance in the extended
space of Floquet system, and (iv) develop a generalization of the
non-Abelian Berry matrix to Floquet system.

\subsection{Derivation of Eq. (7) in the main text}

We want to calculate matrix elements of the effective Floquet
Hamiltonian in a basis of the rotating frame~\cite{eckardt05}. The
starting point is Eq.~(6) in the main text:
\begin{eqnarray}
 \hat{H}_{\rm Floquet}&=&\langle\langle \{n_i\};m |\hat{H}(t)-i\partial_t |\{n_i'\};m'  \rangle\rangle \nonumber\\
 & =& \frac{1}{T}\int_{0}^{T} dt \langle\{n_i\}|
 e^{\frac{iK\sin(\omega t)}{\omega}\sum_{i} (\hat{n}_{i\uparrow}+\hat{n}_{i\downarrow})}\;
   \left(\hat{H}_0 + m\omega \right)
\; e^{-\frac{iK\sin(\omega t)}{\omega}\sum_{i}
(\hat{n}_{i\uparrow}+\hat{n}_{i\downarrow})}
    |\{n_i'\}  \rangle e^{-i(m-m')\omega t}.
\end{eqnarray}
where $\hat{H}_{0}$ is shown in Eq.~(3) in the main text. We note
that the operator $\sum_{i}
(\hat{n}_{i\uparrow}+\hat{n}_{i\downarrow})$ fails to commute only
with the superconducting term
\begin{equation}
\hat{H}_{\rm SC} =
\Delta\sum_{i}\left(\hat{c}_{i\uparrow}^{\dagger}\hat{c}_{i\downarrow}^{\dagger}+h.c.\right).
\end{equation}
Therefore, we have
\begin{eqnarray}
 \hat{H}_{\rm Floquet}&=& \delta_{m m'}\big[\langle\{n_i\}|
\hat{H}_{0}(\Delta=0)|\{n_i'\}  \rangle +m\omega \big]\nonumber\\
& & + \frac{1}{T}\int_{0}^{T} dt \langle\{n_i\}|
 e^{\frac{iK\sin(\omega t)}{\omega}\sum_{i} (\hat{n}_{i\uparrow}+\hat{n}_{i\downarrow})}\; \hat{H}_{\rm SC}\;
 e^{-\frac{iK\sin(\omega t)}{\omega}\sum_{i} (\hat{n}_{i\uparrow}+\hat{n}_{i\downarrow})}
    |\{n_i'\}  \rangle e^{-i(m-m')\omega t}\nonumber\\
&=& \delta_{m m'}\big[\langle\{n_i\}|
\hat{H}_{0}(\Delta=0)|\{n_i'\}  \rangle +m\omega \big]\nonumber\\
& & + \frac{1}{T}\int_{0}^{T} dt \langle\{n_i\}| \hat{H}_{\rm SC} +
\Big( i \frac{2K\sin(\omega t)}{\omega} \Big)
 \hat{H}_{\rm SC}^{AH} + \frac{1}{2!}
 \Big( i \frac{2K\sin(\omega t)}{\omega} \Big)^2 \hat{H}_{\rm SC}+
 \cdots\cdots  |\{n_i'\}  \rangle e^{-i(m-m')\omega t}\nonumber\\
&=& \delta_{m m'}\big[\langle\{n_i\}|
\hat{H}_{0}(\Delta=0)|\{n_i'\}  \rangle +m\omega \big]\nonumber\\
& & + \frac{1}{T}\int_{0}^{T} dt \langle\{n_i\}|\Big(  \hat{H}_{\rm
SC} \cos\left[\frac{2K}{\omega}\sin(\omega t) \right]
 + i \hat{H}_{\rm SC}^{AH} \sin\left[\frac{2K}{\omega}\sin(\omega t) \right]
 \Big)|\{n_i'\}  \rangle e^{-i(m-m')\omega t},
\end{eqnarray}
where $\hat{H}_{\rm SC}^{AH}$ is an anti-Hermitian operator
\begin{equation}
\hat{H}_{\rm SC}^{AH} =
\Delta\sum_{i}\left(\hat{c}_{i\uparrow}^{\dagger}\hat{c}_{i\downarrow}^{\dagger}-h.c.\right).
\end{equation}
Since different photon sectors (labeled by $m$) are separated by the
energy $\omega$, then in the limit $\omega\gg D$ ($D$ is the
band-width), we need consider only the zero photon sector, $m=0$.
This is the rotating wave approximation (RWA). Then, the effective
Floquet Hamiltonian becomes
\begin{equation}
 \hat{H}_{\rm Floquet}= \langle\{n_i\}|
 \hat{H}_{0}(\Delta=0) + J_{0}(2K/\omega)\hat{H}_{\rm SC} |\{n_i'\}  \rangle
\end{equation}
Here, $J_{0}(x)$ denotes zero order Bessel Function of the first
type. The effective Floquet Hamiltonian within RWA has exactly the
same form as $\hat{H}_{0}$ in Eq. (3) in the main text except that
the pairing coupling $\Delta$ is renormalized to $\Delta_{\rm
eff}=\Delta\,J_{0}(2K/\omega)$. Note that the integral of the second
term, $i\hat{H}_{SC}^{AH}$, is zero for $m=m'$. As we emphasized in
the text, non-diagonal terms, $m\neq m'$, proportional to
$J_{m-m'}$, may be important. We have given one example for the case
of Zeeman field modulation.

\subsection{More data for Quasi-energy spectrum}

In the main text we showed a few representative examples of the
quasi-energy spectrum as a function of driving period or amplitude.
In this section we elaborate on this analysis and uncover a rather
complicated structure of the quasi-energy states. This spectrum was
found by numerical diagonalization keeping all the non-diagonal
terms in the effective Floquet Hamiltonian.

\begin{figure}[t]
\centering \vspace{0.1in}
\includegraphics[width=4.5in,clip]{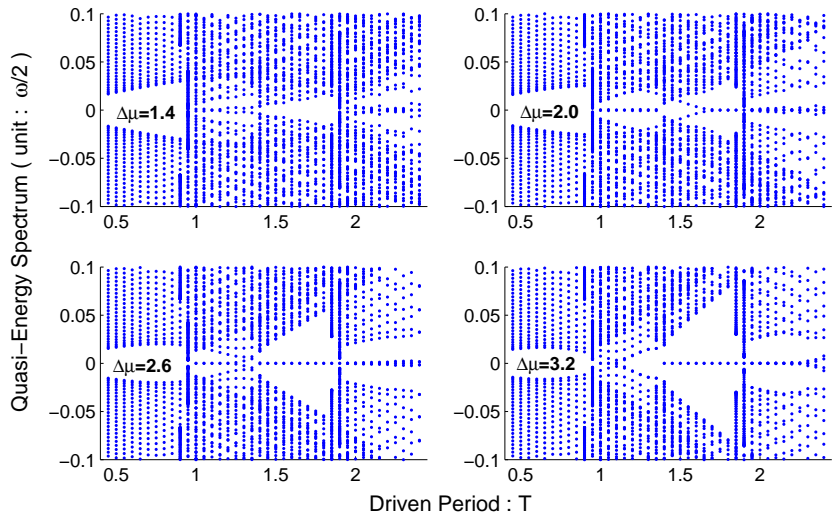}
\caption{(color online) Quasi-energy spectrum ear $\epsilon=0$ for
square-wave driven chemical potential as a function of driving
period $T$. Parameters: $\eta=1.5$ (full band-width $D=4\eta=6.0$),
$V_z=1.0$, $\Delta=1.0$, $\lambda_{\rm SO}=1.2$, and
$(\mu_1+\mu_2)/2=0.5$. Upper panel: $\Delta\mu=|\mu_1-\mu_2|=1.4$
(left), $\Delta\mu=2.0$ (right). Lower panel: $\Delta\mu=2.6$
(left), $\Delta\mu=3.2$ (right). There are 260 sites in the chain.
Note: the energy unit for the quasi-energies is $\omega/2=\pi/T$. }
\label{fig:DrivenMu_supp}
\end{figure}

Fig.\ref{fig:DrivenMu_supp} shows the quasi-energy spectrum as a
function of driving period $T$ for square-wave chemical potential
driving. As the driving amplitude increases from $\Delta\mu=1.4$ to
$2.0$, the Floquet MF in the region $T\in (1.4,1.9)$ appears
gradually, and the quasi-energy gap becomes larger for the FMF in
the region $T\in (1.0,1.35)$. However, as the driving amplitude
further increases, the quasi-energy gap becomes smaller (from
$\Delta\mu=2.0$ to $2.6$) for FMFs in the region $T\in (1.0,1.35)$.
The $T\in (1.0,1.35)$ FMFs are even be killed for $\Delta\mu=3.2$.

\begin{figure}[b]
\centering \vspace{0.25in}
\includegraphics[width=4.5in,clip]{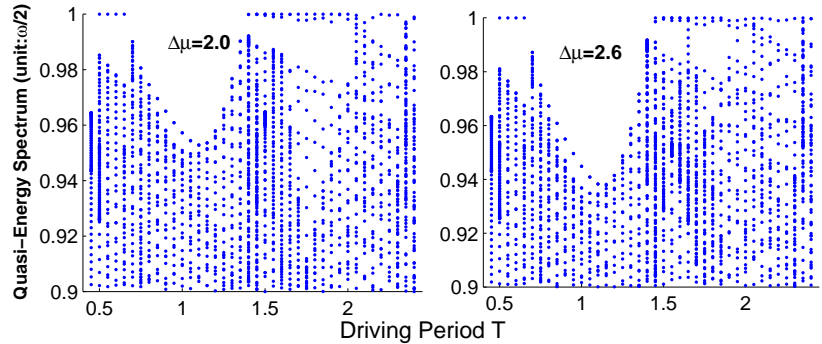}
\caption{(color online) Quasi-energy spectrum near
$\epsilon=\omega/2$ for square-wave driving chemical potential.
Parameters: $\eta=1.5$ (full band-width $D=4\eta=6.0$), $V_z=1.0$,
$\Delta=1.0$, $\lambda_{\rm SO}=1.2$, and $(\mu_1+\mu_2)/2=0.5$. The
spectrum is as a function of driving period $T$ for
$\Delta\mu=|\mu_1-\mu_2|=2.0$ (left), and for  $\Delta\mu=2.6$
(right). $N=260$. } \label{fig:DrivenMu_EW_supp}
\end{figure}

Fig. \ref{fig:DrivenMu_EW_supp} shows the the quasi-energy spectrum
near $\epsilon=\omega/2$ with periodic chemical potential as a
function of driving period $T$. Clearly, one can see the
$\epsilon=\omega/2$ FMF mode appears in large region of parameter
space. The quasi-energy spectrum near $\epsilon=\omega/2$ with
periodic Zeeman splitting is shown in
Fig.~\ref{fig:DrivenMF_EW_supp}, which also reveals FMFs.

\begin{figure}[htp]
\centering \vspace{0.25in}
\includegraphics[width=4.5in,clip]{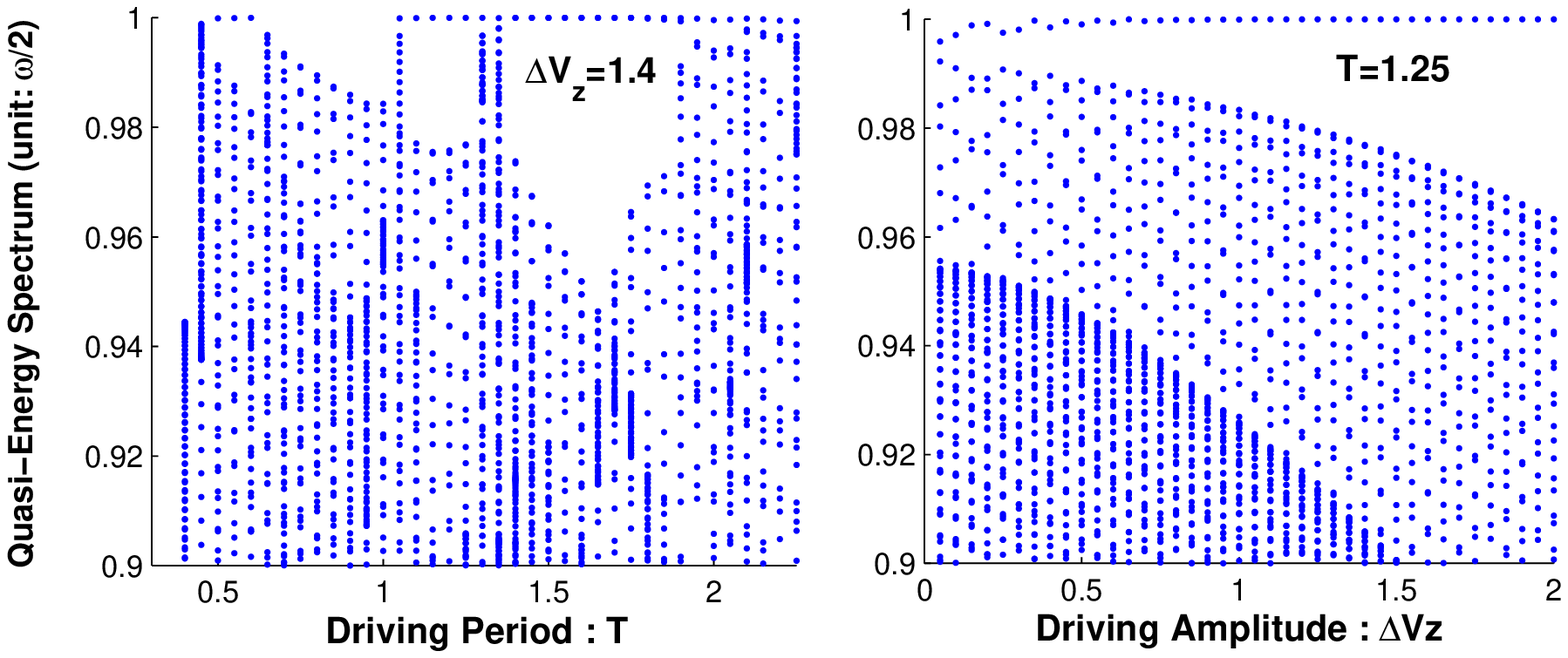}
\vspace{-0.2in} \caption{(color online) Quasi-energy spectrum near
$\epsilon=\omega/2$ for square-wave driving of the Zeeman splitting.
Parameters: $\eta=1.4$ (full band-width $D=4\eta=5.6$), $V_z=1.0$,
$\Delta=2.0$, $\lambda_{\rm SO}=1.5$, and $(\mu_1+\mu_2)/2=0.0$.
$N=260$. Left panel: quasi-energy as a function of driving period
$T$ for $\Delta V_z=1.4$. Right panel: quasi-energy as a function of
driving amplitude $\Delta V_z$ for $T=1.25$. }
\label{fig:DrivenMF_EW_supp}
\end{figure}

\begin{figure}[htb]
\centering \vspace{0.1in}
\includegraphics[width=4.0in,clip]{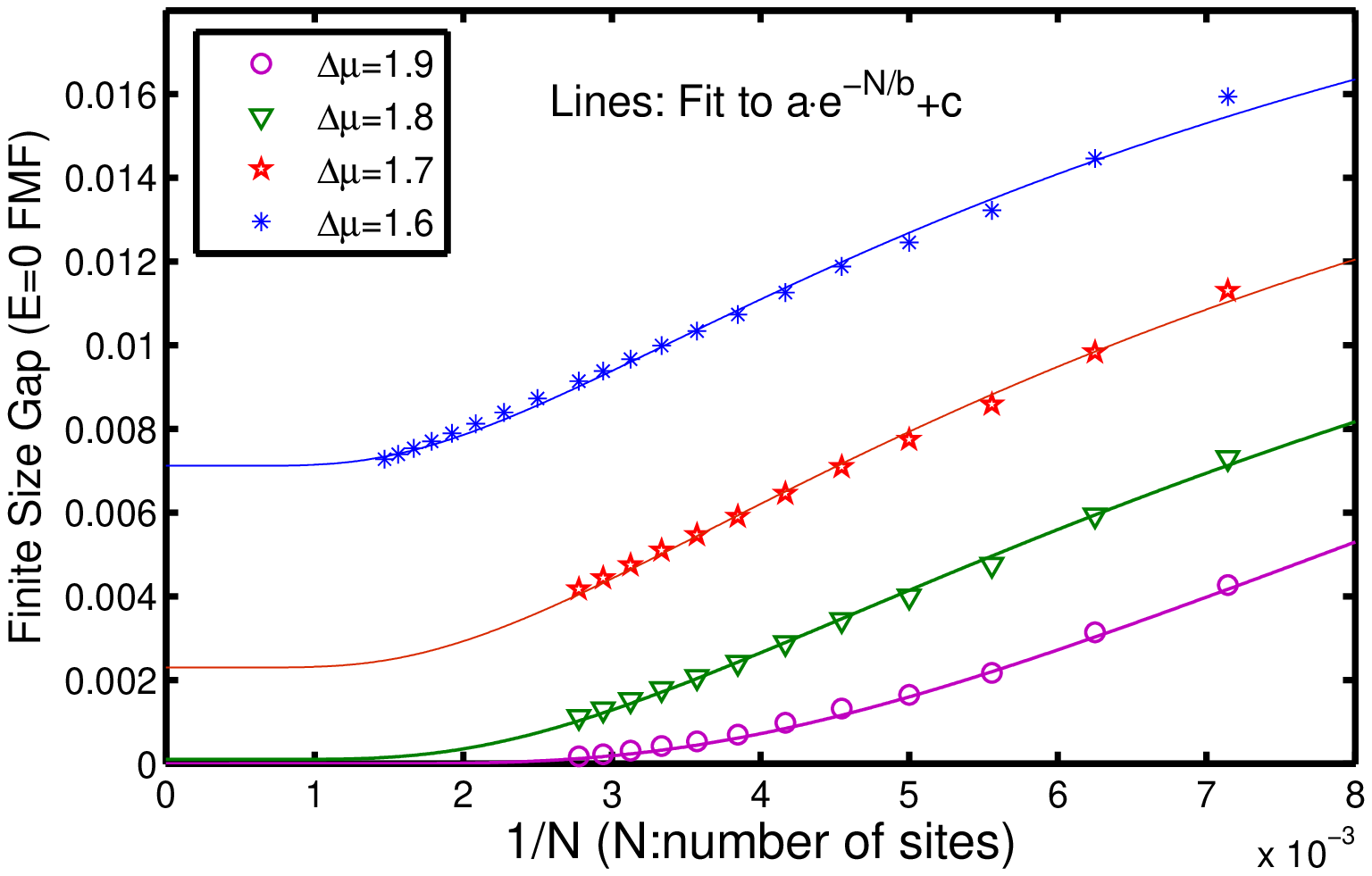}
\caption{ (color online) Finite size splitting (indicating the
coupling between two FMFs of two ends) for $\epsilon=0$ mode as a
function of $1/N$, where $N$ indicates the number of sites in the
chain. Parameters: $\eta=1.5$ (full band-width $D=4\eta=6.0$),
$V_z=1.0$, $\Delta=1.0$, $\lambda_{\rm SO}=1.2$, and
$(\mu_1+\mu_2)/2=0.5$. The period is $T=1.75$. The curves from
bottom to top are for $\Delta\mu=1.9, 1.8, 1.7, 1.6$. The lines show
the best fit to the function $a\cdot\exp(-N/b)+c$. }
\label{fig:FiniteSize_T175}
\end{figure}

Fig. \ref{fig:FiniteSize_T175} shows finite size splitting, which
indicates the coupling between two FMFs at wire ends, as a function
of $1/N$ for the $\epsilon=0$ mode. $N$ is the number of sites in
the chain. The splitting of two FMFs should decay exponentially as
the length of the wire increase: $\sim \exp(-L/\xi)$, where the
length of the wire $L=a N$ with lattice spacing $a$, and $\xi$ is
the superconducting coherent length. We fit the data using the
function form $a\cdot\exp(-N/b)+c$. As $N\rightarrow\infty$, the
finite size splitting goes to zero for $\Delta\mu=1.9, 1.8$, and
goes to finite values for $\Delta\mu=1.7, 1.6$, which shows the
threshold $\Delta\mu_{c}$ for the FMF is larger than $1.7$ and
smaller than $1.8$.

One important quantity in Floquet theory is the mean energy
\cite{fainshtein78} corresponding to the average expectation value
of Hamiltonian over a full period: $E_{\alpha}=(1/T)\int_0^T dt
\langle\phi_{\alpha}(t)|
\hat{H}(t)|\phi_{\alpha}(t)\rangle=\epsilon_{\alpha}-\omega \partial
\epsilon_{\alpha}/\partial\omega$. Numerical data shown indicate
that the partial derivative part of this expression vanishes for MF
zero modes (except near the transition point), so
$E_{0}=\epsilon_0=0$ and
$E_{\omega/2}=\epsilon_{\omega/2}=\omega/2$. Once weak heating
effects are considered, one expects that the driven system tends to
occupy the Floquet state with the lowest mean energy
\cite{arimondo12,ketzmerick10}. Therefore, we focus on the
$\epsilon=0$ FMF, that have the lowest mean energy.

\subsection{$U(1)$ gauge invariance in the extended space of Floquet
system}

For the system described by the BdG Hamiltonian, there is a $U(1)$
gauge invariance \cite{ivanov01}: if the global phase of the
superconducting order parameter $\Delta$ is shifted by $\phi$, i.e.
$\Delta\rightarrow\Delta e^{i\phi}$, it is equivalent to rotating
electron creation and annihilation operator $\psi\rightarrow
e^{i\phi/2}\psi$ and $\psi^{\dagger}\rightarrow
e^{-i\phi/2}\psi^{\dagger}$.

Let us consider the instantaneous quasi-energy eigenstates (also
shown in Eq. (9) in the main text)
\begin{equation}
 \left[\hat{H}(R(\tau),t)-i\partial_t  \right]
 |\phi_{\alpha}(R(\tau),t)\rangle =
 \epsilon_{\alpha}(R(\tau))|\phi_{\alpha}(R(\tau),t)\rangle.
\end{equation}
The quasi-energy state of the Floquet system is defined in a
extended Hilbert space \cite{sambe73}, therefore, one can expand the
operator $\hat{H}(R(\tau),t)$ and state $
|\phi_{\alpha}(R(\tau),t)\rangle$ as the sum of different
photon-sector \cite{sambe73}
\begin{eqnarray}
 \hat{H}(R(\tau),t) &=& \sum_{n} e^{-i n\omega t} \hat{H}_{n}(R(\tau)),\\
  |\phi_{\alpha}(R(\tau),t)\rangle &=&  \sum_{n} e^{-i n\omega t} |\phi_{\alpha n}(R(\tau))\rangle .
\end{eqnarray}
Then, the Floquet Hamiltonian of the extended Hilbert space can be
written as
\begin{equation}
 \hat{\textbf{H}}=
  \begin{pmatrix}
   \ddots & \vdots & \vdots &\vdots&  \\
   \cdots &\hat{H}_{0}+\omega & \hat{H}_{1} & \hat{H}_{2} &\cdots \\
   \cdots &\hat{H}_{-1} & \hat{H}_{0} & \hat{H}_{1} &\cdots \\
  \cdots &\hat{H}_{-2} & \hat{H}_{-1} & \hat{H}_{0}-\omega &\cdots \\
      & \vdots &\vdots&\vdots&\ddots
  \end{pmatrix} \;.
\end{equation}
It is easy to check that the $U(1)$ invariance exists for all the
matrix elements:
$\cdots\hat{H}_{-2},\hat{H}_{-1},\hat{H}_{0},\hat{H}_{1},\hat{H}_{2},\cdots$.
Therefore, the $U(1)$ gauge invariance also exists for the Floquet
system defined in the extended Hilbert space.

The Floquet MF excitations shown in Eq.(10) in the main text can
also be written as
\begin{equation}
 \hat{\gamma}_{\alpha}^{\dagger}(t) = \sum_{n} e^{-i n\omega t} \hat{\gamma}_{\alpha n}^{\dagger},
\end{equation}
where
\begin{equation}
 \hat{\gamma}_{\alpha n}^{\dagger} = \int d\textbf{r} \big[ u_{n}(\textbf{r},R(\tau))\hat{\psi}^{\dagger}(\textbf{r})
     +  v_{n}(\textbf{r},R(\tau))\hat{\psi}(\textbf{r}) \big].
\end{equation}
When the over all phase of the superconducting order parameter
change $2\pi$, the Floquet MF excitation change sign:
$\gamma_{\alpha n}^{\dagger}\rightarrow -\gamma_{\alpha
n}^{\dagger}$ and thus $\gamma_{\alpha}^{\dagger}(t)\rightarrow
-\gamma_{\alpha}^{\dagger}(t)$.

\subsection{Non-Abelian Berry Matrix for Floquet System}

As shown in Eq.(8) in the main text, the adiabatic evolution of the
Floquet system can be described by a two-time Schr\"{o}dinger
equation \cite{breuer89,althorpe97} (also see Eq.(8) in the main
text)
\begin{equation}
 i\partial_{\tau}|\Phi(R(\tau),t)\rangle =
 \left[\hat{H}(R(\tau),t)-i\partial_t  \right] |\Phi(R(\tau),t)\rangle,
\label{eq:twotimeE}
\end{equation}
Here, we will show if the quasi-energy degeneracy occurs in Floquet
system, the adiabatic evolution of the Floquet system can be
described by a generalized non-Abelian Berry matrix, as in the
static counterpart  \cite{wilczek&zee}.

Consider an instantaneous quasi-energy equation for a $k$-fold
degenerate Floquet states
\begin{equation}
 \hat{H}(R(\tau),t)|\phi_{\alpha}(R(\tau),t)\rangle = \epsilon(\tau) |\phi_{\alpha}(R(\tau),t)\rangle\, ,
\end{equation}
where $\alpha=1,2,\cdots k$. If the quasi-energy difference is large
between this subspace and other states, the transitions to the
states outside the $k$-fold subspace can be neglected within the
adiabatic approximation. For any time $\tau$, the wave function of
the system can be decomposed into a linear combination of the
Floquet quasi-energy state
\begin{equation}
 |\Psi (R(\tau),t)\rangle = e^{-i\int_{0}^{\tau}\epsilon(\tau')d\tau'}
\sum_{\alpha} c_{\alpha}(\tau) |\phi_{\alpha}(R(\tau),t)\rangle
\label{eq:expansion}
\end{equation}
Feed Eq.~(\ref{eq:expansion}) into Eq.~(\ref{eq:twotimeE}), one can
obtain
\begin{equation}
 \sum_{\alpha}\Big(i\partial_{\tau}c_{\alpha}(\tau)\Big)|\phi_{\alpha}(R(\tau),t)\rangle
  + \sum_{\alpha} c_{\alpha}(\tau) i\partial_{\tau} |\phi_{\alpha}(R(\tau),t)\rangle  = 0\; ,
\end{equation}
Projecting the equation to the state $\langle
\phi_{\beta}(R(\tau),t)|$ and carrying out the integral
$(1/T)\int_{0}^{T}dt$, one finds
\begin{equation}
 i\partial_{\tau}c_{\beta}(\tau)=-\sum_{\alpha} M_{\beta\alpha} c_{\alpha}(\tau)
\end{equation}
where
\begin{equation}
 M_{\beta\alpha}(\tau)=\frac{1}{T}\int_{0}^{T}
 dt \langle \phi_{\beta}(R(\tau),t)|i\partial_{\tau}  |\phi_{\alpha}(R(\tau),t)\rangle
\end{equation}
corresponds to elements of the $k$-by-$k$ matrix $\textbf{M}(\tau)$.
Then, it is easy to check if the system is initially in the Floquet
state $|\phi_{\alpha}(R(\tau=0),t)\rangle$, the time ($\tau$)
evolution of such state can be written as
\begin{equation}
 |\Psi_{\alpha}(R(\tau),t)\rangle =
 \hat{\textbf{U}}(\tau) |\phi_{\alpha}(R(\tau=0),t)\rangle,
\end{equation}
where
\begin{equation}
 \hat{\textbf{U}}(\tau) = \mathbb{P}\exp\left[ i\int_0^{\tau} \textbf{M}(\tau') d\tau' \right]
\end{equation}
and $\mathbb{P}$ denotes the path-ordering. This is the evolution
operator given in Eq.(12) of the main text.

\end{widetext}

\begin{thebibliography}{99}
\bibitem{note}
The chemical potential in Ref.~\cite{Lutchyn10,oreg10} is
$\mu\to\mu+2\eta$.
\end{thebibliography}
\end{document}